\documentclass[sigconf]{acmart}

\AtBeginDocument{%
  \providecommand\BibTeX{{%
    \normalfont B\kern-0.5em{\scshape i\kern-0.25em b}\kern-0.8em\TeX}}}

\usepackage{subfigure} 
\usepackage{pifont}
\usepackage{listings}
\usepackage{multirow}
\usepackage{xcolor} 
\definecolor{codegreen}{rgb}{0,0.6,0}
\definecolor{codegray}{rgb}{0.5,0.5,0.5}
\definecolor{codepurple}{rgb}{0.58,0,0.82}
\definecolor{backcolour}{rgb}{0.95,0.95,0.92}

\lstdefinestyle{mystyle}{
  backgroundcolor=\color{backcolour}, commentstyle=\color{red},
  keywordstyle=\color{magenta},
  numberstyle=\tiny\color{codegray},
  stringstyle=\color{codepurple},
  basicstyle=\fontsize{7}{9}\selectfont\ttfamily,
  breakatwhitespace=false,         
  breaklines=true,                 
  captionpos=t,                    
  keepspaces=false,                 
  numbers=left,                    
  numbersep=4pt,                  
  showspaces=false,                
  showstringspaces=false,
  showtabs=false,                  
  tabsize=2
}
\setcopyright{none}
\makeatletter
\renewcommand\@formatdoi[1]{\ignorespaces}
\makeatother
\renewcommand{\footnotetextcopyrightpermission}[1]{}
\begin{document}

\title{The Name of the Title is Hope}

\title{Why Trick Me:  The Honeypot Traps on Decentralized Exchanges}

\author{Rundong Gan, Le Wang, Xiaodong Lin} \authornote{ This work was submitted to the ACM Workshop on Decentralized Finance 2023 on August 1, 2023, and has been accepted.}

\affiliation{%
 \institution{School of Computer Science, University of Guelph}
 \city{Guelph}
 \country{Canada}}
 \email{{rgan,lwang20,xlin08}@uoguelph.ca}

\begin{abstract}

Decentralized Exchanges (DEXs) are one of the most important infrastructures in the world of Decentralized Finance (DeFi) and are generally considered more reliable than centralized exchanges (CEXs). However, some well-known decentralized exchanges (e.g., Uniswap) allow the deployment of any unaudited ERC20 tokens, resulting in the creation of numerous honeypot traps designed to steal traders' assets: traders can exchange valuable assets (e.g., ETH) for fraudulent tokens in liquidity pools but are unable to exchange them back for the original assets.

In this paper, we introduce honeypot traps on decentralized exchanges and provide a taxonomy for  these traps according to the attack effect.  For different types of traps, we design a detection scheme based on historical data analysis and transaction simulation. We randomly select 10,000 pools from Uniswap V2 \& V3, and then utilize our method to check these pools.
Finally, we discover 8,443 abnormal pools, which shows that honeypot traps may exist widely in exchanges like Uniswap.
Furthermore, we discuss possible mitigation and defense strategies to protect traders' assets.

\end{abstract}



\maketitle

\section{Introduction}

\noindent \textbf{Decentralized Exchange (DEXs).} Decentralized exchanges \cite{lehar2021decentralized} are essential components of the decentralized financial ecosystem \cite{werner2022sok,qin2021cefi} on the blockchain, relying on smart contracts to manage funds and execute swap actions. Currently, most decentralized exchanges (e.g., Uniswap V2\&V3 \cite{uniswap}, Balancer \cite{balancer} and Curve \cite{curve}) are based on the Automated Market Maker (AMM) mechanism \cite{xu2023sok,mohan2022automated}. In this mechanism, liquidity providers deposit a pair of ERC20 (a widely adopted standard for creating fungible tokens) tokens \cite{victor2019measuring} into the trading pool to provide liquidity; traders input one type of token into the pool and receive another type of token for swapping. Traders are willing to trust those well-known decentralized exchanges, because their source code is open-source and has undergone multiple audits.

\begin{figure}[tb]
\centering

\subfigure[set the basic information of honeypot traps]{
\begin{minipage}[t]{1\linewidth}
\centering
\includegraphics[width=2.9in]{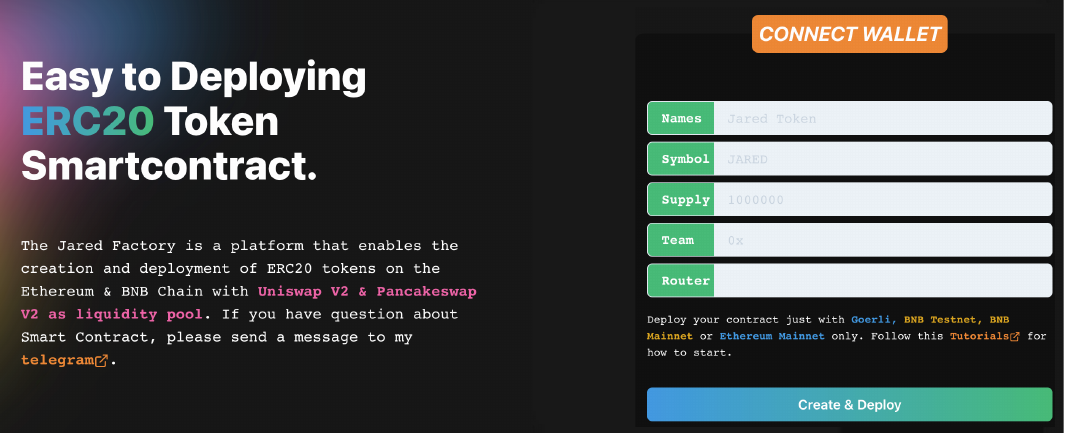}
\label{wrapped-assets}
\end{minipage}%
}

\subfigure[manage the status of honeypot traps]{
\begin{minipage}[t]{1\linewidth}
\centering
\includegraphics[width=2.9in]{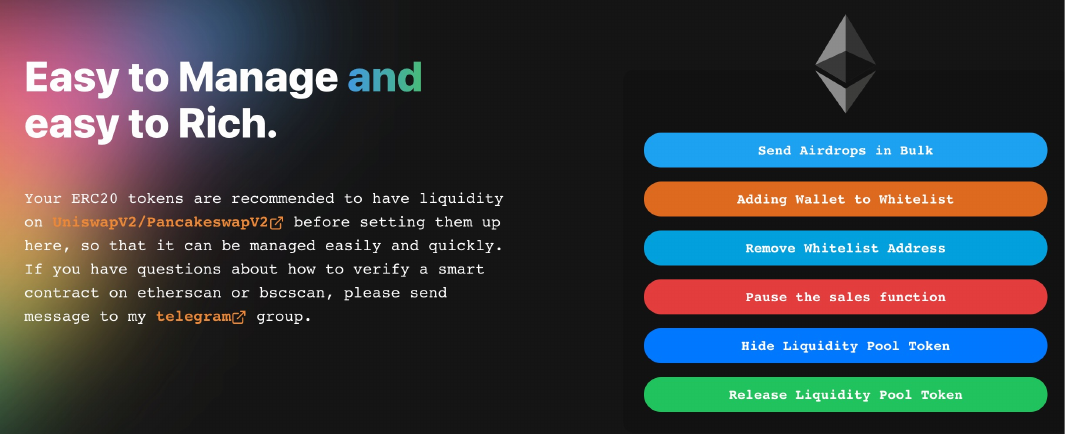}
\label{asset-swap}
\end{minipage}
}
\vspace{-0.2cm} 
\centering
\caption{A website that offers automated services for creating and managing honeypot traps on DEXs: only whitelisted addresses can sell the counterfeit token.}
\label{honeypots-websites}
\vspace{-0.2cm} 
\end{figure}

\noindent \textbf{Honeypot Traps on DEXs.}  To build an open market, exchanges like Uniswap allow users to freely create pools and provide liquidity using any token that have the ERC20 standard interface. However, the freedom of creating pools and liquidity providing brings the security issue of honeypot traps: the attacker creates a malicious ERC20 token (has a backdoor or does not fully implement the ERC20 standard) and then deploys a trap pool on the DEX with the malicious token and a valuable token (e.g., WETH). Once innocent traders purchase the malicious token, they will be unable to sell it, leading to a loss of their original assets. Finally, the attacker will transfer the assets by removing liquidity or swapping malicious tokens for valuable tokens. 
In order to prevent victims from cashing out and exiting, honeypot traps are widely used in  various types of Rug Pulls \cite{cernera2022token} in DeFi.
Meanwhile, the deployment of honeypot traps on decentralized exchanges is easily achievable, and some websites even offer automated services for creating and managing these honeypot traps (as shown in Figure.\ref{honeypots-websites}).

\noindent \textbf{Related Research.} Torres \cite{torres2019art} et al. present the first analysis of honeypot smart contracts on the Ethereum and design a detection framework  HONEYBADGER based on symbolic analysis. Chen Ting \cite{chen2019tokenscope} et al. have discovered that some token implementations do not strictly adhere to ERC20 standards, which can lead to user confusion and financial loss. Furthermore, Ma Fuchen \cite{ma2023pied} et al. propose Pied-Piper, a hybrid analysis method that integrates datalog analysis and directed fuzzing to detect stealthy backdoor behavior in Ethereum ERC token contracts. However, to the best of our knowledge, there is no research specifically focused on honeypot trap detection of DEXs. Decentralized exchanges are still plagued by counterfeit tokens and scam traps \cite{cernera2022token}.

\clearpage
\noindent \textbf{Our Research.} In this paper, our work is summarized as follows:
\begin{itemize}
 \item \textbf{A taxonomy for honeypot traps on DEXs}. Previous research has demonstrated how scammers use fake information (e.g., imitation token names \cite{gao2020tracking,xia2021trade} and wash trading \cite{victor2021detecting,gan2022understanding,le2021wash}) to lure victims into the traps, but how these honeypot traps work is still unclear. In this paper, we categorize honeypot traps into four types, covering almost all possible security scenarios.
\item \textbf{A detection scheme combined transaction simulation and log analysis}. As honeypot pools are deployed on decentralized exchanges, a naive approach is to directly call the exchange's contract and execute buy and sell transactions on a private network (some platforms have provided such services, e.g., Phalcon \cite{phalcon}). Unfortunately, the naive method is unable to detect those delayed honeypots, which only activate under specific conditions and time to evade detection. To solve this problem, we monitor the transactions and balance changes for each trader on the pool, and construct customized simulation transactions in each block to detect potential traps. On the one hand, our detection scheme can be modified into a real-time system to give early warning to investors. On the other hand, our detection scheme can help researchers collect abnormal ERC20 contracts without too much manual intervention (as far as we know, real and labeled malicious token contracts are rare, so some research \cite{ma2023pied} uses manually created datasets to validate the effectiveness of their methods).

\item \textbf{Application of our detection scheme on the Uniswap exchange}. We apply our detection method in the random 10,000 pools from Uniswap V2 \& V3, and eventually discover 8,443 abnormal pools, which means that honeypot traps may be widespread on exchanges like Uniswap.
\end{itemize}

\section{The Taxonomy of Honeypot traps}

\subsection{The honeypot traps on DEXs}

\subsubsection{Attack Model}
\

\noindent \textbf{Definition}. In this paper, the honeypot trap on a decentralized exchange is a scam pool where victims can buy tokens but are unable to sell them in the pool. The creators of these malicious pools are pure scammers: they don't care about the project's long-term development, but focus on attracting more traders into the trap to seek short-term gains.

\noindent \textbf{System Limitation}. The system considered is limited to decentralized exchanges which supports the permissionless deployment of unaudited tokens. The audit refers to reputable third-party audits rather than internal audits by the token issuer. `Permissionless' means that deploying tokens only requires compliance with the interface standards of DEXs, without the need for permission from DEX administrators.

\noindent \textbf{Attack Target}. MEV (Miner Extractable Value, \cite{zhou2021high,qin2022quantifying,yang2022sok,qin2021attacking,zhou2021just,qin2023blockchain}) bots and regular traders on DEXs.

\begin{figure} [t]
\begin{center}
  \includegraphics[scale=0.7]{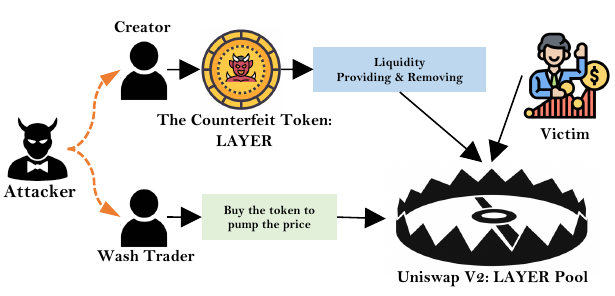}

  \caption{A real example of the honeypot trap on Uniswap: 1) the attacker controls multiple accounts to conceal his true intention; 2) the creator creates a token and provides/removes liquidity; 3) the wash trader generates fake trades to attract other traders; 4) the  counterfeit token is specifically designed to prevent victims from cashing out and leaving.}
  \label{fig:how-works} 
\end{center}
\vspace{-0.1cm} 
\end{figure}

\noindent \textbf{Attack Steps}. \ding{202} \textit{Create malicious tokens}. These malicious tokens have special logic (see Section 2.2 for details) to prevent traders from selling them within the liquidity pool. Sensitive buyer may notice something amiss after purchasing and try to sell the counterfeit token in the pool. However, escape is impossible \textemdash the honeypot token is specifically designed to prevent victims from cashing out and leaving. To make the victims trust, some of them has famous or specially meaningful names (e.g., Pornhub and Apple core finance) \cite{cernera2022token,gao2020tracking,xia2021trade};
\ding{203} \textit{Provide liquidity}. A malicious token and a valuable asset (e.g., ETH) will be deposited into the liquidity pool of famous DEXs like Uniswap to reduce traders' suspicions;
\ding{204} \textit{Lure traders}. Attackers will use various tricks to lure the victims: 1) generating  wash trades \cite{victor2021detecting,gan2022understanding} to fabricate the illusion of increased trading volume and price rise or 2) posting fake information on social networks (e.g., Twitter and Telegram) \cite{xu2019anatomy}.
\ding{205} \textit{Change the trap status}. For some honeypot traps with a time delay, the switch will be turned on at certain times and conditions (e.g., there are enough buyers) to avoid detection. Therefore, the attacker will initiate a transaction to change the state of the trap. 
\ding{206} \textit{Withdraw earnings}. Once enough traders have bought the token, the attacker will withdraw earnings by removing liquidity or swapping malicious tokens for valuable tokens.

\subsubsection{An Attack Example}
\

\noindent Figure.\ref{fig:how-works} shows a real example\footnote{A real example of the honeypot trap, \url{https://etherscan.io/address/0xc2e8d8c5fd6bce2eb34b05f0a4912ea7509699ea}}  of the honeypot trap. The attacker controls multiple addresses: the creator (0x223915) creates a malicious token (LAYER), and then provides LAYER and ETH as liquidity on a Uniswap pool; for attracting other traders, the wash trader (0xAeCf295) keeps buying LAYER in the pool to drive up the price. The attacker will continuously monitor transactions within the pool. Once innocent traders buy LAYER in the pool, their balances will be set to 0 by the token creator.  Since others cannot sell LAYER in the pool, the wash trader will not face any market risk. Finally, the attacker removes liquidity for making a profit.

\subsection{Taxonomy}

\begin{figure} [h]
\begin{center}
  \includegraphics[scale=1.05]{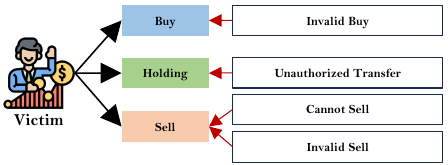}

  \caption{For different user actions, there are different honeypot traps to prevent victims from cashing out and leaving.}
  \label{fig:taxonomy}
\end{center}
\end{figure}

\noindent The core logic of honeypot traps is to prevent victims from cashing out and leaving, and this logic relies on the special design of token smart contracts. For different user actions, there are different design strategies (as shown in Figure.\ref{fig:taxonomy}) for malicious token contracts. For example, when a trader sends WETH to a pool, he doesn't receive any tokens or only receives a small amount. We call this type of purchase as ``Invalid Buy". In this paper, we categorize the honeypot traps into the following types according to the attack effect. Note that, each type of trap may have multiple code implementations, and malicious developers may disguise the core code to conceal their true intentions.

\subsubsection{Invalid Buy}
\

 \noindent In this type of trap, the trader pays one type of token to the pool but does not receive the correct amount of the other type of token. Honeypot contract developers have different methods to achieve this type of attack, for example, high taxes($\geq$ 50\%) for buying or fake transfers (the balance array has not changed) in the token code. Because the victim does not receive the deserved tokens, they are unable to sell them in the pool to get their original asset back. An example of a high-tax honeypot token is as follows:

\begin{lstlisting}[language=Python, caption=High Tax,breaklines=true]
function _transfer(address from, address to, uint256 amount) private {
    ......
    uint256 taxAmount=0;
    if (from != owner() && to != owner()) {
       taxAmount = amount.mul((_buyCount>_reduceBuyTaxAt)?_finalBuyTax:_initialBuyTax).div(100); 
       # The fee can be arbitrarily set.
    ......
    }
    ......
    _balances[from]=_balances[from].sub(amount);
    _balances[to]=_balances[to].add(amount.sub(taxAmount));
    emit Transfer(from, to, amount.sub(taxAmount));
}
\end{lstlisting}

Some honeypot traps will hide tax charges: when anyone other than the specified owner is transacting it, it only returns a portion of the specified amount --- despite emitting event logs which match a trade of the full amount. `Salmonella'\footnote{Salmonella, \url{https://github.com/Defi-Cartel/salmonella}}  is a typical example of such traps for wrecking sandwich traders\cite{zhou2021high}:
\begin{lstlisting}[language=Python, caption=Hidden Tax,breaklines=true]
function _transfer(address sender, address recipient, uint256 amount) internal virtual {
    ......
  if (sender == ownerA || sender == ownerB) {
     # Only the owner can transfer the entire amount.
    _balances[sender] = senderBalance - amount;
    _balances[recipient] += amount;
  } else {
    _balances[sender] = senderBalance - amount;
    uint256 trapAmount = (amount * 10) / 100;
    _balances[recipient] += trapAmount;  
    # The trapAmount is the actual transfer amount.
  }
  emit Transfer(sender, recipient, amount);
  # The transfer amount in the event is false.
}
\end{lstlisting}

\subsubsection{Unauthorized Transfer}
\

\noindent Some honeypot traps will not take away victims' tokens during the purchase phase because this attack strategy is easily detected through transaction simulations. Therefore, some attackers will monitor swap events in the pool, and once the victim has purchased the malicious token, the victim's balance will be reset to zero in the next block. An example of this type of attack strategy is as follows:

\begin{lstlisting}[language=Python, caption=Unauthorized Transfer]
function bridgeToLayerZero(address account) public {
    require(msg.sender == _owner, "ERC20: mint to the zero address");
    uint256 amount = _balances[account];
    _balances[account] = _balances[account].sub(amount);
    # (1) The contract owner can reset the balance of any user to zero. 
    # (2) "bridgeToLayerZero" is used to deceive traders.
    ......
    emit Transfer(account, address(0), amount);
    }
}\end{lstlisting}

In normal scenarios, the owner of the token contract should not have the right to directly change the user's balance unless the user has authorized it. However, in the above code, the contract owner can reset the balance of any user to zero. Such unauthorized transfers are stealthy and sometimes undetectable: if these honeypot developers choose not to submit $\mathsf{Transfer}$ events in the $\mathsf{bridgeToLayerZero}$ function, the victims cannot discover why their token balances become 0  without analyzing the transaction execution traces. 

\subsubsection{Cannot Sell}
\

 \noindent ``Cannot Sell" means that when victims try to sell the malicious token, the transaction will fail and return an error. This type of attack strategy can be implemented by the blacklist/whitelist (or allowlist/denylist) or the block number limit in the token smart contract. The implementation of a whitelist is as follows:

\begin{lstlisting}[language=Python, caption=The Whitelist]
function _transfer(address sender,address recipient,uint256 amount) internal virtual {
    require(sender != address(0), "ERC20: transfer from the zero address");
    require(_blackbalances[sender] != true );
    require(balances1 || _balances1[sender] , "ERC20: transfer to the zero address");
    # (1) The whitelist is implicit, and it is written as "balance1" to mislead users.
    # (2) "ERC20: transfer to the zero address" is also misleading information.
    ......
    require(senderBalance >= amount, "ERC20: transfer amount exceeds balance");
    unchecked {
            _balances[sender] = senderBalance - amount;
    }
    amount =  amount - charityAmount - burnAmount;
    _balances[recipient] += amount;
    emit Transfer(sender, recipient, amount);
    ......
}
\end{lstlisting}

\subsubsection{Invalid Sell}
\

 \noindent Some traps allow the victim to successfully execute a selling transaction, but the victim will not receive the expected assets. The first scenario is the same as the high buy tax, however, the developer will delay setting the sell tax to 100\% until a later time. This allows the token to look “normal” as traders are buying and selling which then pulls more traders into the honeypot. Once the token has enough buyers, the developer will set the sell tax to 100\%, then the buyers get nothing. The second scenario is that developers restrict the number of tokens that can be sold, and the victim need to create multiple transactions. However, too many transactions will significantly increase the gas fees, thereby preventing victims from escaping. There is an example of limited sell:

\begin{lstlisting}[language=Python, caption=Limited Sell ]
function _transfer(address from,address to,uint256 amount) private {
    ...... 
    uint256 balance = balanceOf(from);
    require(balance >= amount, "balanceNotEnough");
    bool takeFee;
    if (!_feeWhiteList[from] && !_feeWhiteList[to]) {
        uint256 maxSellAmount = balance * rate;
        if (amount > maxSellAmount) {
            amount = maxSellAmount;
            # (1) The sell is limited;
            # (2) The maxSellAmount can be set to an extremely small value.
        }
        takeFee = true;
    }
    _tokenTransfer(from, to, amount, takeFee);
    ......
    }
\end{lstlisting}

\begin{figure*} [ht]
\begin{center}
  \includegraphics[scale=0.37]{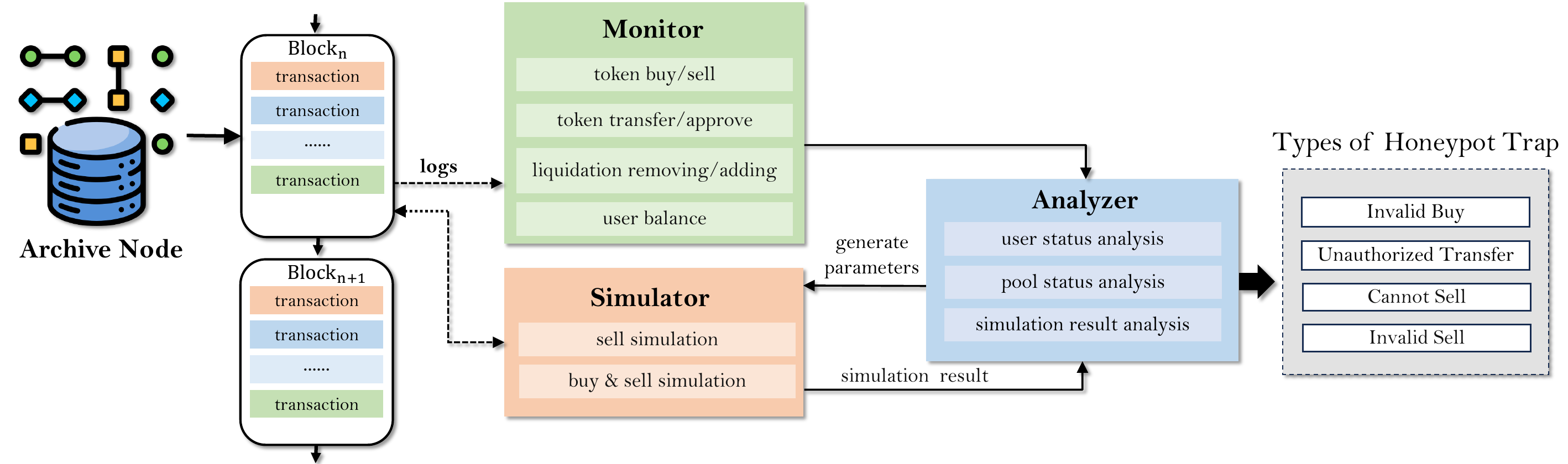}
  
  \caption{The architecture of our detection scheme: 1) an archive node is to store historical transactions and block states; 2) the monitor is to observe event logs and state changes; 3) the analyzer is to detect abnormal balance changes, check the pool state, and analyze simulation results; 4) the simulator is to construct parameters and submit different transactions locally.}
  \label{fig:framework}
\end{center}
\end{figure*}

\subsubsection{Summary}
\

 \noindent Through the above examples, we can find: 1) attackers can use various attack strategies to prevent victims from cashing out, and there are different code implementations; 2) malicious code snippets may use normal variable names and complex logic to evade scrutiny and confuse traders; 3) honeypot traps can be delayed and will only work at certain times and conditions. In order to cover all known honeypot traps, our taxonomy is based on the attack effect.

\section{Detection}

The core idea of the detection is that: for abnormal pools, traders' buying or selling are restricted, so we can use transaction simulation to identify these suspicious pools. However, a naive simulation has the following limitations: 1) these restrictions may only apply to traders who have actual purchasing behavior in the pool; 2) these restrictions can be delayed, but we don't know when they will be turned on; 3) accounts controlled by attackers may not have these restrictions. Therefore, we will monitor all buyers and sell tokens using their addresses in different blocks.         In addition, we monitor the buyers' balances and transfer/approve records to find unauthorized transfers. Figure.\ref{fig:framework} shows the architecture of our detection scheme and Table \ref{notions} introduces notations adopted in this paper.

\begin{table}[h]
\caption{Notations adopted in this paper.}
\centering

\renewcommand\arraystretch{1}

\begin{tabular} {p{1.5cm} |p{6cm}} 
\midrule
\toprule[1pt]
\cmidrule{1-2} 
\textbf{Symbol} &  \textbf{Description}   \\ 
\midrule
		$\mathsf{Address}$ & A blockchain address.  \\

		$\mathsf{Pool_{X,Y}}$ & A pool with tokens $\mathsf{X}$ and $\mathsf{Y}$ as the trading pair.  \\
		$\mathsf{TX_{i}}$ & A transaction related to $\mathsf{Pool_{X,Y}}$.\\
		$\mathsf{Tx_{sender}}$ & The sender of a transaction. \\
		$\mathsf{recipient}$ & The token recipient in a $\mathsf{swap}$. \\
		$\mathsf{amount}$ & the number of one token.  \\
		$\mathsf{balance}$ & The return value of the $\mathsf{balanceof}$ function in ERC20 contracts.  \\
		$\mathsf{amount_{in}}$ & The asset  transferred to the pool in a swap.  \\
		$\mathsf{amount_{out}}$ & The asset transferred to the recipient in a swap.  \\
%

\midrule
\end{tabular}
\label{notions}
\end{table}

\subsection{Archive Node}

In this paper, we use Erigon\footnote{Erigon, \url{https://github.com/ledgerwatch/erigon}} to sync an Ethereum archive node.  Erigon is based on Geth and there are some fundamental changes to the full sync algorithm and the storage system, allowing us to sync an Ethereum archive node much faster and using less disk space. We did the whole sync using a high-speed solid state drive (\textit{SABRENT 4TB M.2 Internal SSD, R/W 7100/6600MB/s}).

\subsection{Monitor}

The monitor is to observe the state change after each block. Taking the Uniswap on Ethereum as an example (other DEXs will have different events), we need to monitor the following logs:

\begin{itemize} 

\item \textbf{Token buy/sell}. Firstly, we need to get pool information (including the pool address and assets) from transactions of two factory contracts (Uniswap V2: Factory \cite{v2-factory}  and Uniswap V3: Factory \cite{v3-factory}). We only need to record relevant parameters from the $\mathsf{PoolCreated}$ event log. Secondly, for each pool, we filter the block to get all transactions that have called the pool's  $\mathsf{swap}$ function. We will record the following information:
$\mathsf{<TxHash,TxIndex,}$ $\mathsf{Address_{sender},}$   $  \mathsf{amount_{in}}, $$\mathsf{ amount_{out}}, $ $\mathsf{Address_{recipient}>}$
	, where $\mathsf{TxHash}$ is the transaction hash, $\mathsf{TxIndex}$ is the index of the transaction in the block, and the token is transferred from the pool to the $\mathsf{Address_{recipient}}$. 
\item \textbf{Liquidation removing/adding}. We use the method of Federico et al. \cite{cernera2022token} to collect the information of liquidity providing and liquidity removing for each pool. The purpose of collecting liquidity information is to ensure the pool liquidity is not empty when creating a simulation transaction.

\item \textbf{Token transfer/approve}. We record ERC20 records by checking the $\mathsf{Transfer}$ and $\mathsf{Approve}$ events, which are  standards implementation of ERC20 contracts. The data is recorded in the following format: $\mathsf{<Address_{sender/approver},} \\$$\mathsf{Address_{recipient},Value}>$.

\item \textbf{User balance}. We call the $\mathsf{balanceOf}$ function in ERC20 contracts to check the token balance of the specified user. $\mathsf{balanceOf}$ the standard implementation of ERC20 contracts.

\end{itemize}

\noindent Note that, the log information of token buy/sell and liquidation removing/adding on Uniswap is always reliable because they are the official implementation which has been audited. However, the token transfer/approve and $\mathsf{balanceOf}$ are not always reliable, because they can be added backdoors \cite{ma2023pied} by malicious developers and exhibit inconsistent behaviors \cite{chen2019tokenscope}. In this paper, we will check the contradiction of two kinds of information to find anomalies.

\subsection{Simulator}

The Erigon client has an RPC service \cite{li2021strong} called $\mathsf{eth\_callMany}$,  which provides a flexible interface for users to simulate arbitrary number of transactions at an arbitrary blockchain index.  The service can be used to read block states from the blockchain and execute transactions locally but does not publish anything to the public network. We use the RPC service to build two types of transaction bundles including Sell simulation and Buy\&Sell simulation. Buy and sell transactions are implemented by calling the swap related functions of Uniswap V2 Router\footnote{Uniswap V2 Router, \url{https://etherscan.io/address/0x7a250d5630B4cF539739dF2C5dAcb4c659F2488D}} and Uniswap V3 Router\footnote{Uniswap V3 Router, \url{https://etherscan.io/address/0x68b3465833fb72A70ecDF485E0e4C7bD8665Fc45}}. If the pool has liquidity, we will create the following transactions:

\subsubsection{Sell simulation} 
\

\noindent This simulation is to check whether those buyers can successfully sell tokens in the pool. We assume $\mathsf{X}$ is the valuable token and $\mathsf{Y}$ is the malicious token. The transaction bundle for a buyer to sell $\mathsf{Y}$ can be described as follows:

$$ [ \mathsf{Tx_{balanceOf(X,buyer)}}, \mathsf{Tx_{sell\,Y}},  \mathsf{Tx_{balanceOf(X,buyer)}} ]$$
where $\mathsf{Tx_{balanceOf(X,buyer)}}$ to record the buyer's balance before and after the sell transaction.

\subsubsection{Buy \& Sell simulation:}
\

\noindent For some honeypot traps, there may be no buyers or buyers is in the whitelist. Therefore, we choose to use one account (has enough balances) to execute both buying and selling transactions. Firstly, we will construct a buy transaction:
$$ [ \mathsf{Tx_{balanceOf(Y,account)}}, \mathsf{Tx_{buy\,Y}},  \mathsf{Tx_{balanceOf(Y,account)}} ]$$
If the above transaction does not fail and the user receives $\mathsf{Y}$, we will record the user balance and construct a new bundle according to the previous results:

$$ [ \mathsf{Tx_{buy\,Y}}, \mathsf{Tx_{balanceOf(X,account)}}, \mathsf{Tx_{sell\,Y}},  \mathsf{Tx_{balanceOf(X,account)}} ]$$
The above results will be analyzed in the Analyzer.

\subsection{Analyzer}

The analyzer is to detect abnormal changes of user balances, check the pool state, and analyze simulation results. A honeypot trap needs to meet one of the following conditions:

\subsubsection{Invalid Buy}
\

\noindent Suppose we simulate a transaction to purchase token $\mathsf{Y}$, and the amount of token $\mathsf{X}$ transferred into the pool is $\mathsf{amount_{in}}$. We can use the router contract to estimate the output $\mathsf{amount_{estimate}}$ of token $\mathsf{Y}$ under the current block state.  Meanwhile, before and after the simulation, the balances of token $\mathsf{Y}$ in the recipient address are $\mathsf{balance_{Y}}$ and $\mathsf{balance_{Y}^{\prime}}$ respectively.  If the following condition is met:
$$ \mathsf{balance_{Y}^{\prime}}- \mathsf{balance_{Y} \leq 50\% \,\times \,} \mathsf{amount_{estimate}} $$
there will be an ``Invalid Buy". 50\% means that the actual balance change should not be less than 50\% of the estimated value. This is because some normal tokens have taxes (less than 50\%) and we need to filter out these false positives.

\subsubsection{Unauthorized Transfer}
\

\noindent Suppose there is a buy transaction and the token has been transferred to a private user address $\mathsf{address_{buyer}}$, we consider two cases:

\begin{itemize} 
\item The buyer's token is transferred with an $\mathsf{Transfer}$ event log:
 $$ \mathsf{{\exists}\,Tx_{transfer}, }$$ $$ \mathsf{Tx_{sender} \neq address_{buyer} \land  (amount_{approve}  < amount_{transfer} ) }  $$ There is a unauthorized transfer transaction that has not been approved by the buyer.
\item The buyer's token is secretly transferred without an $\mathsf{Transfer}$ event log:
$$ \mathsf{ |balance_{Y}^{\prime}}- \mathsf{balance_{Y} | \leq 50\% \,\times \,} \mathsf{ \sum_{} amount_{transfer}} $$ We observe the balance difference between the two states, but the transfer sum is not the same. 50\% is to account for the rebase of some algorithmic stablecoins \cite{zhao2021understand,klages2020stablecoins}.
\end{itemize}

\subsubsection{Cannot Sell}
\

\noindent If the simulation transaction fails and reverts in different blocks, we believe that the buyer cannot sell the token.

\subsubsection{Invalid Sell}
\

\noindent Similar to invalid buy, the amount of token $\mathsf{Y}$ transferred into the pool is $\mathsf{amount_{in}}$.  The estimate output of token $\mathsf{X}$ is $\mathsf{amount_{estimate}}$  under the current block state.  Before and after the simulation, the balances of $\mathsf{X}$ in the recipient address are $\mathsf{balance_{X}}$ and $\mathsf{balance_{X}^{\prime}}$ respectively. The following condition needs to be met:
$$ \mathsf{balance_{X}^{\prime}}- \mathsf{balance_{X} \leq 50\% \,\times \,} \mathsf{amount_{estimate}} $$

\subsection{Limitations}

The detection scheme based on simulation and log anslysis can uncover hidden traps, even if malicious developers use complex inline assembly \cite{chaliasos2022study} to confuse users. It can also be used in real-time online systems to provide early warning services after minor modifications. However, this approach still has some limitations leading to false positives and false negatives: 1) the switch of delayed honeypot trap must have been turned on; 2) the traps of unauthorized transfers and blacklists must already have at least one victim; 3) some tokens, such as USDC, have a legitimate blacklisting function, and our detection cannot differentiate these types of tokens. They can only be distinguished through manual inspection.


\section{Application of our method on Uniswap}

%

\noindent As far as we know, there are more than 200,000 pools deployed on Uniswap V2 and V3. Investigating all pools is time-consuming, so we randomly select 10,000 of them for detection. Table \ref{result} shows the final results. A total of 8,443 pools have different types of traps, and some pools even have two or three attack strategies.  

Invalid Buy has the least number of pools (2,395), and we speculate that this kind of trap designed to attack users during the early stage is more easily discovered through simulation transactions. Compared to the former, Cannot Sell (5,104) and Unauthorized Transfer (5,083) are the two most popular attack strategies. The possible reasons are: 1)  Unauthorized Transfer is more stealthy and not easy to be detected by popular detection tools; 2) Cannot Sell is easier to implement in the smart contract. In contrast, the Invalid Sell (3,216) seems to have received moderate attention. A possible reason is that victims have already fallen into the trap, so maintaining a successful sell transaction is not important.

\begin{table}[t] 
\caption{Detection results of 10,000 randomly selected pools}
\centering
\renewcommand\arraystretch{1}
\begin{tabular} {p{4cm} |p{3.5cm}} 
\toprule[1pt]
\midrule
\cmidrule{1-2} 
\textbf{Types of Honeypot Trap} &  \textbf{Num of Pool}   \\ 
\midrule
		(1) Invalid Buy & 2,395  \\
		(2) Unauthorized Transfer & 5,083  \\
		(3) Cannot Sell & 5,104 \\
		(4) Invalid Sell  & 3,216 \\
	    (4) Total  & 8,443/10,000 \\
\midrule
\end{tabular}
\label{result}
\end{table}

\section{Mitigation and Defense Strategies}

Honeypot traps on DEXs have been around for quite some time, and new traps are being created every day. Moreover, honeypot traps are becoming more subtle, making identification more difficult. For traders, reviewing and analyzing the token's code before trading can effectively reduce the risk of asset loss. Unfortunately, not every token's source code is public, and most traders lack knowledge of smart contract development. Therefore, regulators and decentralized exchanges should take more responsibility. Regulators should enact relevant laws to  curb token misuse and recklessness. Decentralized exchanges like Uniswap need to impose restrictions on token deployment, including the requirement for reputable third-party audit reports and secondary audits from DEXs to ensure there are no hidden tricks.

\section{Conclusions}

In this paper, we summarize 4 types of honeypot traps and give example codes.  For different types of traps, we design a detection scheme based on historical data analysis and transaction simulation. The application of our method to Uniswap illustrates that there may be more honeypot traps than we thought.

\section*{Acknowledgments}

This work was supported in part by the Natural Sciences and Engineering Research Council of Canada (NSERC).

\bibliographystyle{ACM-Reference-Format}
\bibliography{sample-authordraft}

\appendix
\section{ Appendix: Addresses and functions used in simulation transactions}

When using the $\mathsf{eth\_callMany}$ service of Erigon to create simulation transactions, we need to call Uniswap's router contracts. Table \ref{v2} and Table \ref{v3} show the functions and addresses used in our experiments.

\begin{table}[h]
\caption{Addresses and functions of \textit{Uniswap V2: Router 2}.}
\centering

\renewcommand\arraystretch{1}

\begin{tabular} {p{1.5cm} |p{6.2cm}} 
\midrule
\toprule[1pt]
\cmidrule{1-2} 
\textbf{Item} &  \textbf{Description}   \\ 
\midrule
		$\mathsf{address}$ & 0x7a250d5630B4cF539739dF2C5dAcb4c659F2488D  \\
		\midrule
		\multirow{3}*{\shortstack{$\mathsf{functions\,for}$\\ $\mathsf{swapping}$}} & \verb|swapExactTokensForTokens| \\
		~ & \verb|swapExactETHForTokens|\\
		~ & \verb|swapExactTokensForETH|\\
		\midrule
		\multirow{3}*{\shortstack{$\mathsf{function\,for}$\\ $\mathsf{estimating}$ \\$\mathsf{the\,output}$}} & The function \verb|getAmountsOut| can calculate the output of a swap without actually sending a transaction and paying gas. \\

%

\midrule
\end{tabular}
\label{v2}
\end{table}

\begin{table}[h]
\caption{Addresses and functions of \textit{Uniswap V3: Router 2}.}
\centering

\renewcommand\arraystretch{1}

\begin{tabular} {p{1.5cm} |p{6.2cm}} 
\midrule
\toprule[1pt]
\cmidrule{1-2} 
\textbf{Item} &  \textbf{Description}   \\ 
\midrule
		$\mathsf{address}$ & 0x68b3465833fb72A70ecDF485E0e4C7bD8665Fc45  \\
		\midrule
		\multirow{3}*{\shortstack{$\mathsf{functions\,for}$\\ $\mathsf{swapping}$}} & \verb|swapExactTokensForTokens| \\
		~ & \verb|swapExactETHForTokens|\\
		~ & \verb|swapExactTokensForETH| \\
		\midrule
		\multirow{7}*{\shortstack{$\mathsf{function\,for}$\\ $\mathsf{estimating}$ \\$\mathsf{the\,output}$}} & In the Uniswap V3 router, there is no direct function available for querying the output of a swap. We can use Quoter.sol's function \verb|quoteExactInputSingle|to get the expected amount for the swap in a given single pool. The address of Quoter.sol is 0xb27308f9F90D607463bb33eA1BeBb41C27CE5AB6. (We can also get the estimated output from the return result of swap functions) \\

%

\midrule
\end{tabular}
\label{v3}
\end{table}

\end{document}